# Compact Approximation of Lattice Functions with Applications to Large–Alphabet Text Search

Paolo Boldi      Sebastiano Vigna
Dipartimento di Scienze dell'Informazione
Università degli Studi di Milano, Italy

**Abstract**

We propose a very simple randomised data structure that stores an approximation from above of a lattice-valued function. Computing the function value requires a constant number of steps, and the error probability can be balanced with space usage, much like in Bloom filters. The structure is particularly well suited for functions that are bottom on most of their domain. We then show how to use our methods to store in a compact way the bad-character shift function for variants of the Boyer-Moore text search algorithms. As a result, we obtain practical implementations of these algorithms that can be used with large alphabets, such as Unicode collation elements, with a small setup time.[1].

## 1   Introduction

One of the fastest known algorithms for searching a pattern (i.e., a string) in large texts is the Boyer–Moore algorithm [6], along with its many variants (e.g., [1, 8, 7, 12, 13]). All variants are based on the following simple idea: Let $p$ be the pattern of length $P$ and $t$ be the text of length $T$ to be searched[2]. The pattern occurs in position $k$ if $p_i = t_{k+i}$ for all $0 \leq i < P$. Now, when examining a candidate position $k$, we compare the characters $t_{k+m-1}$, $t_{k+m-2}$, and so on, with the corresponding characters of the pattern. If a mismatch is detected at some point, we have to increment $k$, and consider a new candidate position. To choose the increment for $k$ (called the *shift*), we can exploit many heuristics based on information obtained during the scan, and on the structure of the pattern.

A well-known heuristic used in the basic version of the algorithm (and in almost all variants) is the so-called *bad-character shift*. Suppose that while examining position $k$ we find a mismatch at index $j$, that is, $p_j \neq t_{k+j}$ but $p_h = t_{k+h}$ for $h > j$. Instead of incrementing $k$ by just one, we may want to align the pattern so that the last occurrence of $t_{k+j}$ in the pattern is in position $k + j$, at least in case the last occurrence of $t_{k+j}$ happens before position $j$. Note that if $t_{k+j}$ does not occur at all in the pattern we can directly shift the pattern by $j + 1$.

For instance, if we have a mismatch on the first check, that is, in position $P - 1$, and $t_{k+P-1}$ does not appear in the pattern we can shift the pattern by $P$ (albeit apparently infrequent, on large alphabets and small patterns this case occurs fairly often). This will clearly speed up the search significantly.

---

[1]The ideas described in this paper have been implemented as free software under the GNU General Public License within the MG4J project (http://mg4j.dsi.unimi.it/)

[2]We will write $p_i$ and $t_i$ for the $i$-th character of $p$ and $t$, starting from 0; the characters are drawn from a fixed alphabet $A$.

Thus, in general, the shift is computed as $j - \ell(t_{k+j})$, where $\ell(c)$ denotes the index of the last occurrence of character $c$ in the pattern, or $-1$ in case the character does not appear. If the resulting value is not positive, the pattern is shifted by one position.

The key ingredient for implementing the bad-character shift heuristic is a *shift table* that stores $\ell(c)$. Storing such an information is often overlooked, as it is considered a constant-space component of the algorithm: a vector of $|A|$ integers is indeed sufficient. However, the recent need for internationalisation has made large alphabets such as Unicode [14] and ISO 10646 popular (we are thinking, in particular, of the Java programming language, which uses Unicode as native string alphabet). With these alphabets, storing $\ell$ in a table is out of question (ISO 10646 has space for $2^{31}$ characters).

Obvious solutions come to mind: for instance, storing this information in a hash table. However, this approach raises still more questions: unless one is ready to handle rehashing, it is difficult to estimate the right table size, as it depends on the number of *distinct* characters in the string (even an approximate evaluation would not completely avoid the need for rehashing). Moreover, the table should contain not only the shifts, but also the keys, that is, the characters, and this would result in a major increase of space occupancy. Finally, the preprocessing phase could have a severe impact on the behaviour of the algorithm, in particular on short texts. These considerations hold *a fortiori* for more sophisticated data structures, such as balanced binary trees.

A very simple solution to this problem has been proposed in [9]: observe that we can content ourselves to store upper bounds for $\ell(c)$: this could slow down the search, but certainly will not produce incorrect results. Thus, instead of using a hash table with standard collision resolution, one might simply use a fixed-size table and combine colliding values using maximisation.

This approach, however, has several drawbacks: first of all, it does not allow trade-offs between space and errors; second, the birthday paradox makes it easy to get collisions, even with relatively large tables; third, the technique is patented, and thus cannot be freely used in academic or open source work.

Our proposal is inspired by the good statistical properties of Bloom filters [4], a technique from the early 70s that has seen recently a revival because of its usefulness in web proxies (see, e.g., [10]). Starting from an approximation of the number of distinct characters in the pattern, we provide a way to store an upper bound to $\ell$ that is tunable so to obtain a desired error probability; with respect to a hash table, one of the main advantages is that a bad estimate for the number of distinct characters can make the approximation worse, but it is otherwise handled gracefully. To be as general as possible, we cast our ideas in the framework of monotone approximation of lattice functions.

The interest of the authors in this subject is due to their involvement in the *Managing Gigabytes for Java* project. The project aims at implementing inverted-index compression techniques for large-scale full-text indexing in Java, and takes its name from the well-known book by Witten, Moffat and Bell [15]. One of the key components of the project is a complete rewrite of the Java string classes [5] with better algorithms and low-impact object creation, which raised the need for a simple and effective implementation of a Boyer–Moore like search algorithm (the standard Java classes operate a trivial exhaustive search). In particular, using hash maps or other sophisticated data structures was immediately ruled out due to the high setup cost, and to the hidden (and potentially very heavy) burden of garbage collection; since the latter runs in time proportional to the number of *alive objects*, incrementing the frequency of garbage collection has a cost that can grow independently of all data contained in the algorithm implementation. This aspect is usually overlooked in the choice for more efficient algorithms, but often it backfires (as the Java string classes show).

## 2 Notation

For each natural number $n \in \mathbf{N}$, we denote with $[n]$ the set $\{0, 1, \ldots, n-1\}$.

Let $L$ be a lattice [3], whose partial order relation is denoted by $\leq$; the greatest lower bound (least upper bound, respectively) of two elements $x, y \in L$ will be denoted by $x \wedge y$ ($x \vee y$, resp.). $L$ is assumed to contain a least element, denoted by $\perp$ (bottom).

Let $\Omega$ be a fixed set. Our purpose is to provide a data structure to represent, or to approximate, functions $\Omega \to L$. Let $f : \Omega \to L$ be a function; its *support* $D(f)$ is the set of all elements of the universe that are not mapped to $\perp$, that is,

$$D(f) = \{ x \in \Omega \mid f(x) \neq \perp \}.$$

In our main application, of course, $\Omega$ is the alphabet and $L$ is the lattice of natural numbers.

## 3 Compact Approximators

**Definition 1** Let $d > 0$ and $m > 0$. A *d-dimensional m-bucket compact approximator* for functions from $\Omega$ to $L$ is given by a sequence of $d$ independent hash functions $h_0, h_1, \ldots, h_{d-1}$ from $\Omega$ to $[m]$, and by a vector $b$ of $m$ values of $L$. The elements of $b$ are called *buckets*.

When using an approximator to store a given function $f : \Omega \to L$, we fill the vector $b$ as follows:

$$b_i = \bigvee_{\exists j < d \; h_j(x) = i} f(x)$$

In other words, the vector $b$ is initially filled with $\perp$. For each $x \in \Omega$ such that $f(x) \neq \perp$, a $d$-dimensional approximator spreads the value of $f(x)$ in the buckets of indices $h_0(x), h_1(x), \ldots, h_{d-1}(x)$; when conflicts arise, the approximator stores the maximum of all colliding values.

Now, the function *induced* by the thus filled approximator is defined by

$$\tilde{f}(x) = \bigwedge_{j < d} b_{h_j(x)}$$

In other words, when reading $f(x)$ from an approximator we look into the places where we previously spread the value and compute the minimum.

The interest of approximators lies in the following (obvious) property:

**Theorem 1** For all $x \in \Omega$, $f(x) \leq \tilde{f}(x)$.

Note that in case $L = \{0, 1\}$ we obtain exactly a Bloom filter (by approximating the characteristic function of a subset of $\Omega$), whereas in case $d = 1$ we obtain the structure described in [9].

As an example, consider the function $f : \Omega = [12] \to \mathbf{N}$ given by $f(1) = 3$, $f(5) = 1$, $f(9) = 2$ and $f(x) = 0$ in all other cases. We let $d = 2$, $b = 6$, $h_0(x) = \lfloor x/2 \rfloor$ and $h_1(x) = 5x \bmod 6$ (these functions have been chosen for exemplification only). In the upper part of Figure 1, one can see how values are mapped and maximised into the buckets; in the lower part, one can see how some of the values are extracted. Note that some values are obtained from a single bucket, because $h_0$ and $h_1$ coincide. The values for which $\tilde{f}(x) > f(x)$ are highlighted.

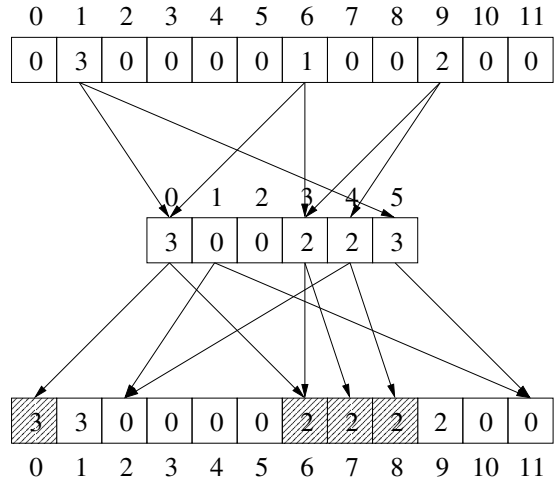

Figure 1: A diagram representing a compact approximator.

Our interest is now in estimating how often $\tilde{f}(x) = f(x)$. We divide our evaluation in two parts: the bottom case and the nonbottom case.

**The bottom case.** Since we have in mind functions with a small support, we should look carefully at
$$\varphi = \Pr[\tilde{f}(x) \neq f(x) \mid f(x) = \bot],$$
that is, the probability of erroneous computation at a point in which $f(x) = \bot$. The analysis is similar to that of a Bloom filter: if $D(f) = n$ (i.e., $f$ is non-$\bot$ in $n$ points), the probability that a bucket contains $\bot$ is
$$\left(1 - \frac{1}{m}\right)^{dn}.$$
To compute the wrong value, we must find non-$\bot$ values in all the buckets over which we minimise. This happens with probability
$$\varphi = \left(1 - \left(1 - \frac{1}{m}\right)^{dn}\right)^d \approx \left(1 - e^{-\frac{dn}{m}}\right)^d$$

The expression on the right is minimised at $d = m \ln 2/n$; the minimum is then $(1/2)^d$. In other words, we can ameliorate exponentially the error probability by using more hash functions and a larger vector; the vector should be sized approximately as $m = 1.44\, dn$.

Note that the presence of multiple hash functions is essential: for instance, when $d = 3$ and $m = 3n/\ln 2$ we have $\varphi = 1/8$, whereas a single hash function gives $\varphi \approx 1/5$.

More precisely, if we set $d = 1$, since
$$1 - e^{-\frac{n}{m}} = \frac{n}{m} + O\left(\frac{n^2}{m^2}\right)$$
when the ratio $m/n$ grows we get inverse linear error decay, as opposed to exponential.

**The nonbottom case.** This is definitely more complicated. Our interest is now in estimating
$$\psi = \Pr[\tilde{f}(x) \neq f(x) \mid f(x) \neq \bot],$$
that is, the probability of erroneous computation at a point in which $f(x) \neq \bot$. Note that
$$\psi = \sum_{i=0}^{s} \Pr[\tilde{f}(x) \neq f(x) \mid f(x) = v_i] \Pr[f(x) = v_i \mid f(x) \neq \bot]$$
$$= \sum_{i=1}^{s} \frac{a_i}{n} \Pr[\tilde{f}(x) \neq f(x) \mid f(x) = v_i]$$

We now need to make some assumptions on the distribution of the values assumed by $f$. Suppose that $f$ assumes values $\bot = v_0 < v_1 < \cdots < v_s$, and that it assumes value $v_i$ exactly $a_i > 0$ times (i.e., $|f^{-1}(v_i)| = a_i$). Then,
$$\Pr[\tilde{f}(x) \neq f(x) \mid f(x) = v_i] = \left(1 - \left(1 - \frac{1}{m}\right)^{d \sum_{j=i+1}^{s} a_j}\right)^d$$
since the event $\tilde{f}(x) \neq f(x)$ takes place iff each of the $d$ buckets assigned to $x$ is occupied by at least one of the $\sum_{j=i+1}^{s} a_j$ elements with values greater than $v_i$. All in all we get
$$\psi = \sum_{i=1}^{s} \frac{a_i}{n} \left(1 - \left(1 - \frac{1}{m}\right)^{d \sum_{j=i+1}^{s} a_j}\right)^d.$$

The previous summation is not going to be very manageable. As a first try, we assume that $f$ takes each value exactly once (as in our main application), getting to
$$\psi = \frac{1}{n} \sum_{i=1}^{n-1} \left(1 - \left(1 - \frac{1}{m}\right)^{d(n-i)}\right)^d,$$
where we used the fact that for $i = s$ the summand is always zero (there is no way to store erroneously the maximum value attained by $f$) and that now $s = n$. Note that for each summand we can apply the usual approximation
$$\left(1 - \left(1 - \frac{1}{m}\right)^{d(n-i)}\right)^d \approx \left(1 - e^{-\frac{d(n-i)}{m}}\right)^d$$

Thus, each summand (which gives $n$ times the error probability for value $v_i$) is minimised by $d = m \ln 2/(n - i)$. This is very reasonable: larger values gain from numerous hash functions, as they are likely to override smaller values. For the smallest value, the probability of error is very close to that of $\bot$. The situation is clearly depicted in Figure 2, which shows the error probability as a function of $d$.

Even if we cannot provide an analytical minimum for $\psi$, the assumption that $D(f)$ is very small w.r.t. $\Omega$ makes a choice of $d$ that minimises $\varphi$ sensible.

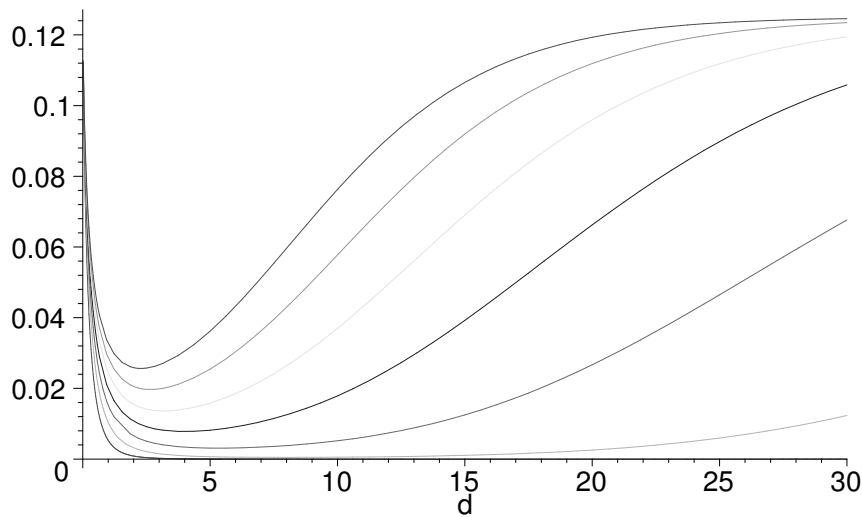

Figure 2: The error probability of each summand in the uniform case, with $n = 8$ and $m = 2n/\ln 2$.

**Exponential Distribution.** A similar partial analysis may be made if the values are distributed exponentially, that is, $v_i = 2^{(s-i)}$; in particular, this means that $f(x) = \bot$ on half of $\Omega$. In this case, $\sum_{j=i+1}^{s} a_j = 2^{s-j+1} - 1$, $n = 2^{s+1} - 1$ and we get

$$\psi = \sum_{i=0}^{s-1} \frac{2^{s-i}}{n} \left(1 - \left(1 - \frac{1}{m}\right)^{d2^{s-i+1}}\right)^d,$$

where this estimate includes (when $i = 0$) the contribution of the bottom case.

In this scenario it is even more sensible to tune the choice of $d$ and $m/n$ using the bottom case. Indeed, all summands behave much better (on one side, the error has a lesser impact as $i$ grows, as $v_i$ decreases exponentially; on the other side, larger values have a greater probability of being stored exactly), as shown in Figure 3.

## 4 Using Approximators in the Boyer–Moore Algorithm

The previous discussion paves our way toward an implementation of the Boyer–Moore algorithm that uses an approximator to store the bad-character shift table. Recall that the function we want to approximate is $\ell : A \to \mathbf{Z}$, where $\ell(c)$ is the index of the last occurrence of $c$ in the pattern, or $-1$ if the character does not occur. For sake of implementation efficiency, we will indeed approximate $\ell' : A \to \mathbf{N}$, with $\ell'(c) = \ell(c) + 1$, so that $\bot = 0$.

Notice that having an upper bound for $\ell'$ is sufficient for the Boyer–Moore algorithm to work correctly, because of the way shifts are computed. More precisely, when analysing a given candidate position $k$ in the text, if the $j$-th character of the pattern is the rightmost mismatch, and $c$ is the text character found in that position (the *bad character*), then we compute the shift as $\max(1, j - \ell'(c) + 1)$. Having a larger value for $\ell'$ has the simple effect of reducing the shift. This is true even for variations of the algorithm that look at different characters to compute the shift (e.g., [13]).

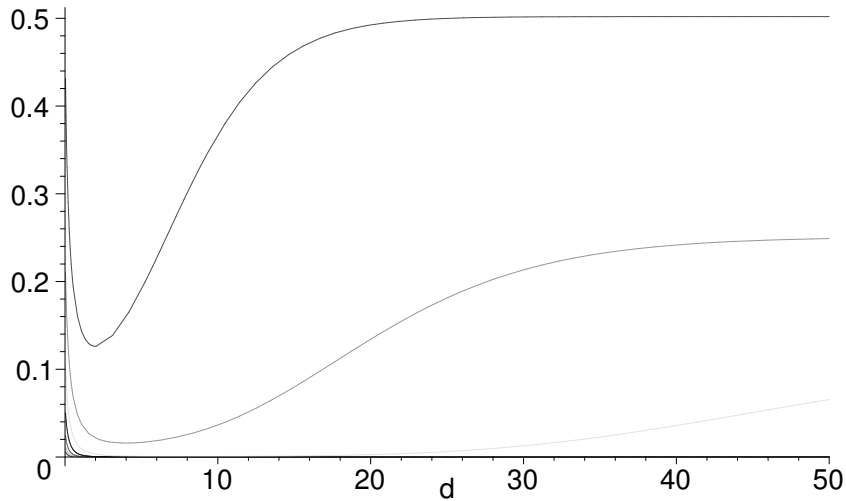

Figure 3: The error probability of each summand in the exponential case with $n = 255$ and $m = 2n/\ln 2$.

## 4.1 Implementation Issues

Suppose we want to find occurrences of pattern $p$ in text $t$. Devising an approximator for $\ell'$ requires choosing the various parameters involved in the approximation.

**Estimating $n$.** As a first step, we have to evaluate the number $n$ of distinct characters in $p$; a rough estimate is given by the length of $p$, but you can try to adopt more sophisticate techniques to get a better bound for $n$ (see, e.g., [11, 2]). Note that these are constant-space, linear-time techniques that give just an approximation, but this is perfectly acceptable, as approximation errors lead only to better or worse precision in representing $\ell'$ (depending on whether the error is on the upper or lower side).

**Choosing $d$ and $m$.** Then, we must decide the number $m$ of buckets and $d$ of hash functions we are going to use for the approximator. As explained above, one should choose $m$ and $d$ so that $d \approx \ln 2\, m/n$.

According to the analysis outlined in the previous sections, a larger value for $d$ reduces the error probability; on the other hand, choosing a value of $d$ that is too large may severely reduce the performance, because both the memory requirement (number of buckets) and the time needed to consult the approximator grow linearly with $d$.

The choice of $d$ and $m$, hence, depends subtly on the quantity of memory available, and on the trade-off between the time one needs to compute the $d$ hash values and the time that is wasted when a short skip is computed as a consequence of the imprecise evaluation of $\ell'$. It is also a good idea to maximise the computed value of $m$ with a reasonable constant, so to compensate the setup overhead on very short patterns.

**The algorithm.** In figure 4 we give a sketch of the algorithm. The function `initialise` chooses the values of $n$, $d$ and $m$, allocates the `bucket` vector and initialises it. The `shift` function computes

```
int bucket[];
int m, n, d;

int hash(char c,int k) { ... }

void initialise(char *p) {
    n=... /* an upper bound for the number of distinct chars in p */
    d=... /* the number of hash functions */
    m=(int)(n*d/ln(2));
    bucket=new array of m integers;

    for (i=0;i<length(p);i++)
       for (k=0;k<d;k++)
          bucket[hash(p[i],k)]=max(bucket[hash(p[i],k)],i+1);
}

int shift(char c,int j) {
    v=MAX_INT;
    for (k=0;k<d;k++)
       v=min(v,bucket[hash(c,k)]);
    return max(1,j-v+1);
}
```

Figure 4: A sketch of the algorithm.

the shift when the bad character is *c* and the position of the mismatch is *j*. Note that the algorithm assumes the presence of a function hash(c,k) that computes the *k*-th hash value of character *c* (hash values are between 0, inclusive, and *m*, exclusive).

**Experimental results.** We implemented the algorithm in Java and run a number of experiments to determine its performance. The most important question concerns the number of positions that are considered for matching; in particular, we are interested in the ratio between the number $c_{app}$ of candidate matching positions considered by our approximate algorithm w.r.t. the optimal number *c* of candidate positions considered by an exact implementation of the Boyer–Moore algorithm using an entire skip table (in both cases, we are only adopting the bad-character heuristic).

Figure 5 shows the ratios $c_{app}/c$ as a function of *d* for various situations. The diagram on the left shows the ratios in the case of three patterns of lengths 9, 18 and 27 made of characters that appear frequently in the text. Clearly, the relative performance of the approximate algorithm decays as the pattern length grows. The diagram on the right shows the same data for patterns made of characters that appear rarely in the text. We remark that in the first case 2 hash functions are sufficient to limit the increment of $c_{app}$ to about 6%, even for the longest pattern. On the contrary, in the second case we need three functions to get similar results. The reason is clear: in the latter case the cost of imprecise representation is high, because almost all characters in the text to be searched should be ⊥-valued (and thus provide a large skip), but this will not happen with a too imprecise representation. The loss is reduced in the former case because some of the most frequent characters are not ⊥-valued. In general, the evaluation of the effectiveness of a compact approximator should be based on a distribution of the inputs that it will process.

The data we gathered suggest that 3 hash functions with buckets $4.3n$ are a good choice; also 2 hash functions with $2.9n$ buckets behave reasonably, but have a sensible loss with rare patterns. Note the strange peak at $d = 5$ for a rare pattern. This phenomenon is almost unavoidable for some values of $d$, and it is due to the fact that increasing the number $d$ of hash functions may cause (in a transient way) an error in the computation of $\ell'$ for a very frequent text character, such as "e" in English. The penalty in this case is very high, because comparisons with that character will be frequent, and will increase substantially the number of positions considered for matching.

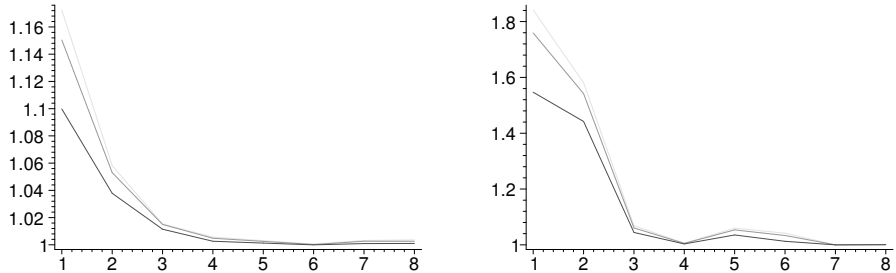

Figure 5: Ratios $c_{app}/c$ for frequent and rare patterns, with varying $d$.

**Benchmarks.** We present benchmarks of the implementation of compact approximators contained in the Java class `TextPattern` provided[3] within the MG4J project (http://mg4j.dsi.unimi.it/). The benchmarks were produced on a Pentium 2.4 GHz running Linux, and on a Sun Fire V880 based on SPARC 900 MHz processors and running Solaris 9, using in both cases the Sun 1.4.1 JDK.

We compare four pattern-search methods:

- a brute-force double loop (as implemented in the `indexOf()` method of `String`);

- an exact implementation of the Boyer–Moore algorithm (more precisely, of its variant known as QuickSearch [13]), using a Java `Map` to store the bad-character shift table[4];

- another exact implementation of QuickSearch, using an array;

- an approximate implementation of the same algorithm, using compact approximators and one of the approximate counting algorithms described in [2].

Our tests were performed on two 16Mbyte documents (one produced at random, and the other one containing a US-ASCII English text), and consisted in searching (all occurrences of) a 9-character and a 54-character pattern. In an attempt to account for garbage-collection overhead, we provide also timings obtained by taking into account also the time required by a call to the `System.gc()` method, which suggests the virtual machine to perform garbage collection. All timings are in milliseconds.

---

[3]It should be noted that the current distribution features several tweaks to fine-tune the techniques presented in this paper: for instance, US-ASCII characters are treated separately using a vector, as they appear frequently in almost every Unicode text.

[4]Since Java does not provide maps handling primitive types without wrappers, we really used a type-specific hash-table map from `fastutil` (http://fastutil.dsi.unimi.it/).

The reader will notice that compact approximators work in all cases much better than maps and better than the brute-force approach (even though, of course, on very short patterns a brute-force loop will outperform any sophisticated algorithm). The timings for arrays are given just for comparison, because, as we already remarked, it is usually not practical to allocate such a large array (a quarter of megabyte in the case of Unicode).

It is interesting to note that, in the case of a long pattern on a random text, the very sparse memory accesses of the array implementation makes it even *slower* than the approximator-based one, as most memory accesses are cache misses.

A final *caveat*: the impact of garbage collection may seem small, but the reader must take into consideration that *almost no objects were alive during the collection*. As we mentioned in the introduction, in a real-world large applications, the collection time may be much larger, even when searching the same pattern within the same text.

|  | Pentium (Linux) | | SPARC (Solaris) | |
| --- | --- | --- | --- | --- |
|  | w/ gc | w/o gc | w/ gc | w/o gc |
| brute force | 60 | | 335 | |
| `Map` | 120 | 114 | 402 | 395 |
| array | 30 | 27 | 190 | 185 |
| approximator | 56 | 52 | 252 | 238 |

Table 1: English 16Mbyte text, 9-character pattern.

|  | Pentium (Linux) | | SPARC (Solaris) | |
| --- | --- | --- | --- | --- |
|  | w/ gc | w/o gc | w/ gc | w/o gc |
| brute force | 55 | | 324 | |
| `Map` | 71 | 65 | 224 | 216 |
| array | 26 | 22 | 113 | 111 |
| approximator | 31 | 28 | 134 | 131 |

Table 2: English 16Mbyte text, 54-character pattern.

|  | Pentium (Linux) | | SPARC (Solaris) | |
| --- | --- | --- | --- | --- |
|  | w/ gc | w/o gc | w/ gc | w/o gc |
| brute force | 50 | | 306 | |
| `Map` | 116 | 108 | 365 | 359 |
| array | 49 | 45 | 189 | 181 |
| approximator | 48 | 42 | 200 | 194 |

Table 3: Random 16Mbyte text, 9-character pattern.

|  | Pentium (Linux) | | SPARC (Solaris) | |
|---|---|---|---|---|
|  | w/ gc | w/o gc | w/ gc | w/o gc |
| brute force | 49 | | 310 | |
| `Map` | 47 | 40 | 144 | 136 |
| array | 37 | 33 | 85 | 83 |
| approximator | 25 | 22 | 92 | 87 |

Table 4: Random 16Mbyte text, 54-character pattern.

## 5 Conclusions

We have introduced a compact approximate representation for lattice functions. This representation is very simple, yet it has good statistical properties and it can be tuned similarly to Bloom filters to obtain specific error bounds. The most interesting aspect of the representation is that, similarly to perfect hashing, it does not need to store elements of the domain. This can be particularly useful if those elements are large or of irregular size, for instance strings.

It is interesting to note that compact approximators are very easily implemented in hardware, as they need no control flow: reading and writing is performed in a fixed number of steps if $m$ and $d$ have been tuned once for all using knowledge about the input data.

[9] Mark E. Davis. Forward and reverse Boyer–Moore string searching of multilingual text having a defined collation order. U.S. Patent 5,440,482, assigned on August 8, 1995 to Taligent, Inc.

[10] Li Fan, Pei Cao, Jussara Almeida, and Andrei Z. Broder. Summary cache: a scalable wide-area Web cache sharing protocol. *IEEE/ACM Transactions on Networking*, 8(3):281–293, 2000.

[11] P. Flajolet and G. N. Martin. Probabilistic counting algorithms for data base applications. *J. Comput. System Sci.*, 31(2), 1985.

[12] Thierry Lecroq. A variation on the Boyer–Moore algorithm. *Theoretical Computer Science*, 92:119–144, 1992.

[13] Daniel M. Sunday. A very fast substring search algorithm. *Comm. ACM*, 33(8):132–142, 1990.

[14] The Unicode Consortium. *The Unicode Standard, Version 3.0*. Addison-Wesley, Reading, MA, USA, 2000.

[15] Ian H. Witten, Alistair Moffat, and Timothy C. Bell. *Managing Gigabytes: Compressing and Indexing Documents and Images*. Morgan Kaufmann Publishers, Los Altos, CA 94022, USA, second edition, 1999.12